\documentclass[12pt]{iopart}

\usepackage{graphicx}
\usepackage{iopams}
\expandafter\let\csname equation*\endcsname\relax
\expandafter\let\csname endequation*\endcsname\relax
\usepackage{mathtools}
\usepackage{dcolumn}
\usepackage{bm}
\usepackage{subfigure}
\usepackage{epsfig}
\usepackage{afterpage}
\usepackage[colorinlistoftodos]{todonotes}
\usepackage[colorlinks=true, allcolors=blue,bookmarks=false,hypertexnames=true]{hyperref}

\newcommand{\ra}{\rangle}
\newcommand{\la}{\langle}
\newcommand{\bea}{\begin{eqnarray}}
\newcommand{\eea}{\end{eqnarray}}
\newcommand{\beq}{\begin{equation}}
\newcommand{\eeq}{\end{equation}}
\newcommand{\be}{\begin{equation}}
\newcommand{\ee}{\end{equation}}
\newcommand{\beqa}{\begin{eqnarray}}
\newcommand{\eeqa}{\end{eqnarray}}

\newcommand{\p}{\partial}

\definecolor{dgreen}{rgb}{0,0.7,0}

\begin{document}
	
\title{Macroscopic fluctuations of a driven tracer in the symmetric exclusion process}


\author{Rahul Dandekar and Kirone Mallick}
\address{ Institut de Physique Theorique, CEA, CNRS, Universite Paris–Saclay, F–91191 Gif-sur-Yvette cedex, France}

\begin{abstract}
	The dynamics of an  asymmetric tracer in the symmetric  simple exclusion process is mapped, in the continuous scaling limit, to the  local current through the origin in the zero-range process (ZRP) with a biased bond. This allows us to study  the hydrodynamics of the SEP with an asymmetric tracer with a step initial condition, leading to the average displacement as a function of the bias and the densities on both sides. We then derive the cumulant generating function of the process in the high-density limit, by using the  Macroscopic Fluctuation Theory and obtain agreement with the microscopic results of Poncet et al (2021). For more general initial conditions, we show that the tracer variance in the high-density limit depends only on the generalized susceptibility in the initial condition.
\end{abstract}


\section{Introduction} \label{sec:introduction}

Stochastic lattice gases of interacting particles are invaluable models to
explore the dynamics of many-body systems:  powerful
theoretical  tools have been developed to study them   in the
physics and the mathematical literature
(some general  references are
 \cite{Spohn91,Zia95,Liggett99,Kipnis99,Chou11}).
 Many of these  processes  appeared in  biophysics to represent 
  molecular motors  or biopolymerization on nucleic acid templates
\cite{MacDonald68,MacDonald69}.  Among  numerous models, 
the one-dimensional simple exclusion  process
-- in which  particles
 perform  random walks subject to the constraint that a site
 can be occupied by at most one particle at a given time --
has acquired a 
paradigmatic status, thanks to its rich  combinatorial structure that
has lead to many exact analytical results \cite{Schutz00,Derrida2007}.
In particular, the macroscopic  hydrodynamics of  the
simple exclusion  process is by now
well understood \cite{Kipnis89,Spohn91} and
 the analysis of the microscopic process allows one to probe 
 fluctuations  beyond the  average hydrodynamic behavior. Indeed, noise at the
 lattice size scale can affect the coarse-grained evolution of the system and
 generate rare events (or  large deviations) with drastic  impact
 \cite{Hugo-rev,Bertini01,Derrida2007}.

 Another manner to explore the interplay between the microscopic and macroscopic scales is to consider a  probe particle (also
 called a  tracer) and follow its  motion in the surrounding fluid of
 particles. If the tracer follows the same dynamical rules as  the other
  particles, it behaves as a passive scalar and  does not affect the 
  hydrodynamics. However, if the probe behaves in a different manner
  (it is sensitive to an external  drive, or it is {\it active} in some way), 
  then the overall collective dynamics can be altered.  This is precisely
  the type of effect we wish to investigate  in the present work.
  
  We shall consider the one-dimensional symmetric simple exclusion process (SEP)
 on the infinite one-dimensional lattice with 
  a single  biased tracer (or tagged particle).  This 
  biased tracer  hops with rate $r_+$ to its right neighbor
  and $r_-$ to its  left, while all other particles hop at rate $1$ in
  either direction. All particles, including the tracer, are  subject to the
  exclusion condition and their order remains  unchanged: this is 
   a single file-dynamics.
  It is known \cite{gleb92,oshanin1999phase,Jussieu13,gleb15} that the average displacement of the  biased tracer in an  infinite  system grows, for large times, as
\beq
\la X_t \ra = \sqrt{2 c t} \label{eq:avgdisp}
\eeq
where $X_t$ is the displacement of the tracer at time $t$, and $c$ is a number that depends on the initial density profile and on the tracer bias through the ratio $\frac{r_+}{r_++r_-}$. 
If the tracer particle is symmetric and has the same dynamics as all other particles,  the variance of the tracer displacement for large times grows as
\beq
\la X_t^2 \ra - \la X_t \ra^2 = \sigma_{\rm{ic}} \sqrt{t} \label{eq:avgvar}
\eeq
where $\sigma_{\rm{ic}}$ is a constant that depends on the initial densities on either side of the tracer and on the initial
set-up of the system   \cite{harris1965diffusion,arratia1983motion, lin2005random,rajesh2001exact,derrida2009current,krapivsky2014large,krapivsky2015tagged}. 
More generally, for  reflecting Brownian motions \cite{krapivsky2014large,hegde2014universal} and  for the SEP \cite{imamura2017large}, it is possible to
calculate  all the cumulants of the position of an unbiased tracer, that  all
grow with time as $\sqrt{t}$.

For  a biased  tracer, the behavior of the variance is not known in most cases even for the SEP. In the high-density limit, $\rho \rightarrow 1$, it has been shown that equation~\eqref{eq:avgvar} is valid for the driven tracer as well, and the coefficient $\sigma_{\rm{ic}}$ can be computed microscopically \cite{Jussieu13,poncet2021cumulant}. In fact, in the high-density limit,
the full cumulant generating function of the tracer position can be calculated
from  microscopic dynamics \cite{poncet2022exact}. However, many  questions
remain unanswered and
a more general framework for studying the driven tracer would be useful.


In the present  work, we  study  the
asymmetric tracer in the framework of  Macroscopic Fluctuation Theory (MFT),
by extending the hydrodynamic approach to asymmetric tracers developed by Landim, Olla and Volchan \cite{landim1998driven,landim2000equilibrium}. For symmetric single-file systems,  the MFT  \cite{Bertini2015, Derrida2007} allows one to derive statistical results for quantities like the current fluctuations \cite{krapivsky2012fluctuations} and the tracer position \cite{krapivsky2015tagged} in terms of the diffusivity and conductivity coefficients of the hydrodynamic description.
However, the  presence
of the biased  tracer generates a moving  boundary condition
at the macroscopic level. A suitable way to derive this condition is to use 
a mapping between single-file systems and mass transfer processes.
In particular, the SEP in 1D is mapped to the 1D Zero-range process (ZRP) \cite{evans2004factorized,evans2005nonequilibrium} and the biased particle in the SEP becomes a biased bond in the ZRP. It can  be shown (see section \ref{sec:driventracer}) that, for any density, the dynamics depends on $r_-$ and $r_+$ only through the parameter $\frac{r_+}{r_+ + r_-}$.
In the 
high-density limit of the SEP, the  MFT  equations can be solved and 
the full cumulant generating function of the tracer position
can be calculated.  Our
results agree  with the microscopic derivations in \cite{Jussieu13,poncet2021cumulant}.
Recently, the dependence of the variance of the tracer position on the initial state has been investigated in a broader class of initial states and shown to depend on the initial state only through a specific quantity, the generalized compressibility \cite{banerjee2022role}. The MFT allows us to study 
these  more general initial states  and we show that
\beq
\la X_t^2 \ra - \la X_t \ra^2 = (\alpha_{\rm{ic}} \sigma_{\rm{annealed}} + (1-\alpha_{\rm{ic}}) \sigma_{\rm{quenched}}) \sqrt{t} \label{eq:avgvar2}
\eeq
where $\alpha_{\rm{ic}}$ is the generalized susceptibility of the initial ensemble, and $\sigma_{\rm{annealed}}$ and $\sigma_{\rm{quenched}}$ are the coefficients of the tracer variance for quenched (deterministic)
and annealed (fluctuating) initial conditions respectively.

The outline of this work  is as follows. In Section \ref{sec:zrpmapping}, we describe the general mapping between single-file systems and a dual mass-transfer process at the hydrodynamic level, that generalizes  
a known microscopic mapping between the SEP and the ZRP. In Section \ref{sec:driventracer}, we first describe the hydrodynamic approach to calculating the average displacement of a driven tracer in the SEP for any density and then, we
derive the complete  MFT boundary conditions at the tracer  using the mapping to a mass transfer process. In section \ref{sec:highdensity}, we solve the case of the SEP in the high-density limit, $\rho \rightarrow 1$ and calculate the full cumulant generating function for the tracer position using the MFT, recovering
the microscopic  results of \cite{Jussieu13,poncet2021cumulant}. We also calculate the dependence of the variance on the initial condition and show that it depends only on the generalized compressibility. Section \ref{sec:conclusions} is devoted to concluding remarks.

\section{Single-file and  mass-transfer processes} \label{sec:zrpmapping}

\begin{figure}[h]
	\centering	
	\includegraphics[width=0.8\linewidth]{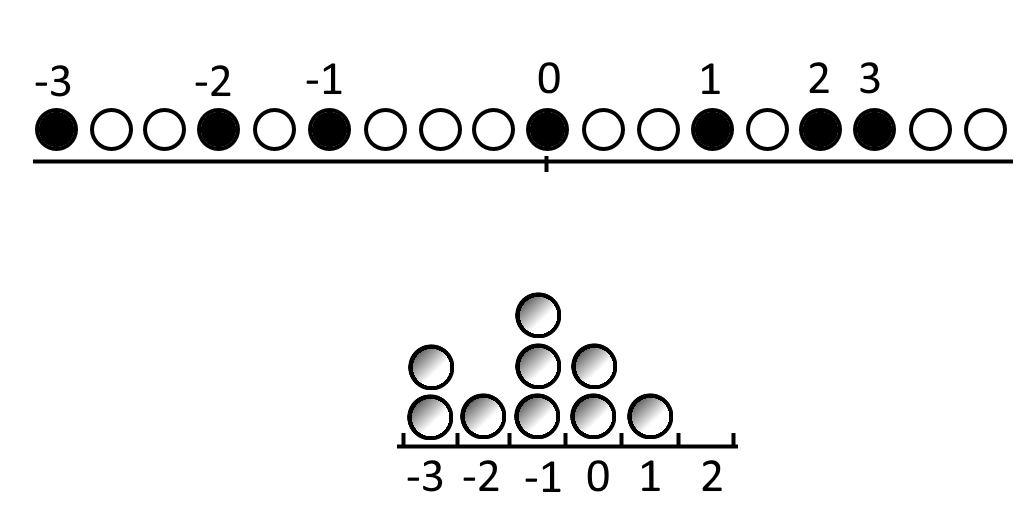}
	\caption{A configuration in the SEP (left) and its corresponding ZRP configuration (right). The  particle with label $i$ in SEP becomes the  site $i$ in ZRP and the number of marbles over  site $i$ in ZRP corresponds to the number of holes between the successive particles $i$ and $i+1$ in SEP.}
	\label{fig:sepzrp}
\end{figure}

In a single-file process on a one-dimensional lattice the particle order is conserved by the dynamical rules, because  overtaking is forbidden.  The  Symmetric Simple Exclusion Process (SEP) is an archetypical
single-file model and its well-known  mapping to the Zero Range Process (ZRP) 
\cite{evans2004factorized,evans2005nonequilibrium} goes as follows (see Fig.~\ref{fig:sepzrp}):
\begin{enumerate}
	\item Label the SEP particles at the initial time ($t=0$), starting from the origin, with $i=0$ being the particle with the smallest non-negative position, $i=1$ being the next particle to the right, and so on. Call $z_i$ the number of empty sites between particles $i$ and $i+1$.
	\item Consider a ZRP with masses $z_i$ on site $i$ and rate $r=1$ of mass transfer from each occupied site to each of its neighbors, irrespective of the mass $z$ on the site.
	\item A mass transfer to the right results in $z_i$ decreasing by one, and $z_{i+1}$ increasing by one. Thus, in the SEP, it corresponds to particle $i+1$ hopping to the left. Thus, the mapping between the SEP and ZRP dynamics is that  the current (or mass transfer)  in the ZRP along the bond $[i-1,i]$ to the left (right) corresponds to the particle with label  $i$ in the SEP
 hopping to the right (left).
\end{enumerate}

The  particle with label $i=0$, having the smallest non-negative initial  position will be hereafter called  the {\it tracer}. Its position with time
will be denoted as $X(t)$. In the ZRP language, the tracer is mapped to the site 0 and tracer movement to mass transfer along the bond $[-1,0]$.

\subsection{Mapping a single-file to a  mass-transfer process}

In this subsection we present a general mapping of the continuum limit of an arbitrary single-file diffusion system to a mass-transfer process, generalizing the mapping of the previous section. Our aim is to relate precisely the statistical fluctuations of these two types  of models. From now on, observables that refer to a single-file process will be written with a  subscript $s$ and the spatial coordinate in the single-file process is denoted by $x$. Quantities related to a  mass-transfer model  have  a subscript $m$ and the spatial coordinate is denoted by $z$. Hence, $\rho_s(x)$ and $\rho_m(z)$ represent the local density in the single-file process and in the  mass-transfer process, respectively.

 Using the same  underlying idea as that the one  explained in the previous subsection,
the transformation laws for the coordinates and the
densities can be written as follows: 
\begin{eqnarray}
  z = \int_{X(t)}^x \rho_s(u,t) \,  du \label{transf:coord} \\
  \int_{0}^z \rho_m(u,t) \,  \rm{d}u   =   \int_{X(t)}^x (1 - \rho_s(u,t)) \,  du
   \label{transf:dens}
  \end{eqnarray}
The first equation says that a  distance $z$ in the mass-transfer process corresponds to  the total number of particles between the tracer (located at $X(t)$) and the position $x$ of the $z^{\rm{th}}$ particle in the single-file.  The second equation expresses the fact that the total mass
between 0 and $z$  in the mass-transfer process is equal to the total number of holes between  $X(t)$ and $x$ in the single-file. Finally,  the position of the tracer  $X(t)$
is given by {\it minus} the mass-transfer  through the origin
\begin{eqnarray}
  X(t) = -J(t) = \int_{0}^\infty  \left(\rho_m(u,0) - \rho_m(u,t)\right) \,  du  
 \label{transf:trace}
  \end{eqnarray}
Equations~(\ref{transf:coord}), 
(\ref{transf:dens}) and (\ref{transf:trace}) define a complete macroscopic one-to-one mapping between the  single-file and the mass-transfer processes. In particular, 
differentiating eqns. \eqref{transf:coord} and \eqref{transf:dens}
with respect to $x$, we deduce that the local densities in the two processes are related through
\begin{eqnarray}
   \rho_s(x,t) = \frac{1}{\rho_m(z,t) + 1} \label{eq:map0}
\end{eqnarray}
The transformation~(\ref{transf:coord}) and (\ref{transf:dens})  can be inverted  as 
\beqa
x(z,t) &=&   X(t) +  z  + \int_0^z  \rho_m(u,t) \,  du,  \label{eq:map1}
\eeqa
and hence
\beqa
\frac{dx}{dz} = 1 + \rho_m(z,t) = \frac{1}{\rho_s(x,t)}.
\eeqa


We now relate the diffusivities of the two models to each other.
Consider the current in a coarse-grained single-file system in the presence of a uniform density gradient,
\beq
j_s = - {D_s}(\rho_s) \frac{d{\rho_s}}{dx} \nonumber
\eeq

This would also cause a current in the mass transfer process. Since transfer across a bond in the mass transfer process corresponds to the motion of a particle (in the opposite direction) in the single-file system, we have, for the current in the mass transfer process,
\beq
j_s = - \rho_s j_m, \mbox{  that is,  } j_{m} =  - \frac{{j_s}}{{\rho_s}}. \label{eq:Jmap}
\eeq

Using  \eqref{eq:map0} and \eqref{eq:map1}, we obtain (note that the density gradient is of opposite sign after the transformation)
\beq
j_m = -\frac{1}{(1+\rho_m)^2}{D_s}\left(\frac{1}{1+\rho_m}\right) \frac{d\rho_m}{dz} = - D_m(\rho_m) \frac{d\rho_m}{dz} \nonumber
\eeq

Hence, the diffusivity for the mass transfer process is related to the diffusivity of the single-file system through the relation
\beq
D_m(\rho_m) = \frac{1}{(1+\rho_m)^2}{D_s}\left( \frac{1}{1+\rho_m}\right), \mbox{  or  } D_s(\rho_s) = \frac{1}{\rho_s^2} D_m\left(\frac{1}{\rho_s}-1 \right)
\label{eq:map2}
\eeq

\subsection{MFT equations and the cumulant generating function for the current}


The Macroscopic Fluctuation Theory is defined in terms of two hydrodynamic coefficients, the diffusivity $D(\rho)$ and the conductivity $\sigma(\rho)$. The Macroscopic Fluctuating Theory equations for a stochastic process with diffusivity $D$ and conductivity $\sigma$ are given, in the bulk,  by
\begin{subequations}
\beqa
\p_t q &=& \p_x(D(q) \p_x q) - \p_x ( \sigma(q) \p_x p) \label{eq:EL1}\\
\p_t p &=& - D(q) \p_{xx} p - \sigma'(q) (\p_x p)^2 \label{eq:EL2}
\eeqa
\end{subequations}

We will consider the calculation of the cumulant generating function for the current across the origin in time $T$, $J(T)$
\beq
\mu(\lambda) = \log{\big\la e^{\lambda J(T)} \big\ra} \label{eq:mudef}
\eeq
where the angular brackets denote the average over all paths, weighted by the path action $S$. $\mu(\lambda)$ can be calculated from the current through the origin in the MFT. (See also \cite{bettelheim2022inverse}). Let the notation $J[\mathcal{P}]$ denote the value of $J$ calculated along a particular trajectory $\mathcal{P}$ of the system. Differentiating, we have
\beq
\mu'(\lambda) = \frac{\la J e^{\lambda J}\ra}{\la e^{\lambda J} \ra} = \la J \ra_{\lambda} \nonumber
\eeq
where the final average is over the tilted action, $S + \lambda J$.\\

Let us assume that, in the hydrodynamic limit, the tilted action is dominated by a single trajectory, $\mathcal{P}_S(\lambda)$, the saddle-point trajectory. Then,
\beq
\mu'(\lambda) = J[\mathcal{P}_S(\lambda)] = J_{MFT}\label{eq:mupformula}
\eeq
since the MFT equations provide the saddle-point solution to the tilted action. One can thus calculate the CGF from the current in the MFT solution, along with the condition that $\mu(\lambda=0) =0$.

The MFT equations \eqref{eq:EL1} and \eqref{eq:EL2} are supplemented by boundary conditions, which depend on the quantity being calculated. For the CGF of the current, the appropriate boundary condition for $p$ at the final time is \cite{derrida2009current}
\beq
p(x,T) = \lambda \theta(x) \label{eq:pft}
\eeq

We start with a step initial density, with uniform density $\rho_{-\infty}$ to the left of the origin and $\rho_{+\infty}$ to the right of the origin, and consider two kinds of initial conditions, annealed and quenched \cite{derrida2009current, Derrida2007,Bertini2015,krapivsky2012fluctuations}. In the quenched case, the step initial condition is deterministic:
\beq
q(x,0) = \rho_{-\infty} \theta(-x) + \rho_{+\infty} \theta(x) \label{eq:qit}
\eeq

In  the annealed case,  the initial profile is allowed to fluctuate,  leading to an implicit equation for $q(x,0)$ in terms of $p(x,0)$:
\beq
p(x,0) = \lambda \theta(x) + \theta(-x) \int_{\rho_{-\infty}}^{q(x,0)} dq \frac{2 D(q)}{\sigma(q)}  + \theta(x) \int_{\rho_{+\infty}}^{q(x,0)} dq \frac{2 D(q)}{\sigma(q)} \label{eq:pit}
\eeq


\subsection{Mapping the MFT equations}

In this section, we show how the conductivity $\sigma(q)$ and the field $p$ transform in the mapping from a single-file process to a mass-transport model. We use the Hamiltonian formulation \cite{Bertini2015} of the MFT equations of motion \eqref{eq:EL1} and \eqref{eq:EL2} in terms of the fields $p_m$ and $q_m$ for the mass-transfer model
\beq
H = \int dz \left( -D_m(q_m) (\p_z q_m) (\p_z p_m) + \frac{\sigma_m(q_m)}{2} (\p_z p_m)^2\right) \label{eq:mfthamil}
\eeq
We transform this  Hamiltonian to  the single-file system
by using equations~\eqref{eq:map0}, \eqref{eq:map1} and \eqref{eq:map2}:
\beqa
H &=& \int dx \left(\frac{dz}{dx}\right)
\bigg( -  q_s^2 {D_s}(q_s) (\p_x q_s^{-1}) (\p_x p_m) \left(\frac{dx}{dz}\right)^2 \nonumber \\
 & &+ \frac{1}{2}{\sigma_m}\left(\frac{1}{q_s}-1 \right) (\p_x p_m)^2 \left(\frac{dx}{dz}\right)^2\bigg) \nonumber\\
&=& \int dx \left( {D_s}(q_s) (\p_x q_s) \left(\frac{\p_x p_m}{q_s}\right)
    +  \frac{q_s}{2}{\sigma_m}\left(\frac{1}{q_s}-1 \right)    \left(\frac{\p_x p_m}{q_s}\right)^2  \right) \nonumber
\eeqa
We demand that this transformation preserves the Hamiltonian structure of the
MFT equations:  hence,  the above  Hamiltonian must take the form
\beqa
H &=& \int dx \left( - {D_s}(q_s) (\p_x q_s)(\p_x p_s) +
\frac{\sigma_s(q_s)}{2} (\p_x p_s)^2 \right) \nonumber
\eeqa
We deduce  the following transformations rules for $(\p_x p)$ and $\sigma$:
\beqa
\p_x {p_s}(x) &=& -\frac{\p_x p_m}{q_s} = - \p_z p_m(z) \\
\sigma_s(\rho_s) &=& \rho_s {\sigma_m}\left(\frac{1}{\rho_s}-1\right), \mbox{  or,  } {\sigma_m}(\rho_m) = ( 1 + \rho_m) \, \sigma_s\left(\frac{1}{1 + \rho_m}\right) \label{eq:sigma-sm}
\eeqa

This completes the mapping at the hydrodynamic level between a single-file system and the corresponding mass transfer process. We can verify that the hydrodynamic coefficients of SEP \cite{Bertini2015} and its corresponding ZRP are related by the transformations found above:
\begin{eqnarray}
D_m(\rho_m) &=& \frac{1}{(1+\rho_m)^2}, ~~~ \sigma(\rho_m) = \frac{2 \rho_m}{1+\rho_m} \label{eq:zrpdiffs} \\
D_s(\rho_s) &=& 1, \quad\quad\quad\quad\quad \sigma(\rho_s) = 2 \rho_s(1-\rho_s)  \label{eq:SEPdiffs}
\end{eqnarray}

\section{Driven tracer in SEP} \label{sec:driventracer}

We now consider a {\it biased} (or driven) tracer in SEP, initially located at the origin, that hops with rate $r_+$ to the right, and rate $r_-$ to the left. The initial density is a step profile  with density $\rho_{-\infty}$
for $x<0$ and density $\rho_{+\infty}$ for $x>0$. 
The particular dynamics of the tracer will affect the density profile in its neighborhood. We start by rederiving the expression of the density profile in a spirit close to that of \cite{gleb92} (see~\cite{AKR09} for  a mathematically similar problem).Then, we shall state the problem in the framework of the MFT and solve it in the high density regime.

\subsection{Density profile and mean displacement}

First we notice that the average density $\rho(x,t)$ of the host particles satisfies the diffusion equation
\begin{equation}
\label{DE}
\frac{\partial \rho}{\partial t} = \frac{\partial^2 \rho}{\partial x^2}
\end{equation}
In the long-time limit,  fluctuations  are ignored and
we treat $X(t)$ as a deterministic variable. The profile  $\rho(x,t)$ as seen from the tracer will adopt a stationary scaling form. This profile  may  be discontinuous at the position of the tracer: we denote the density profile for $x \ge X(t)$ by $\rho_{R}(x,t)$, and the density profile for $x \le X(t)$ by  $\rho_{L}(x,t)$. Similarly, $\rho_{+}$ and $\rho_{-}$ will denote the limiting values of the density just to the right $(x = X(t)+0^+)$ and just to the left $(x = X(t)-0^+)$ of the tracer respectively.

The diffusion equation~(\ref{DE}) has to be supplemented by the following boundary conditions:
\begin{eqnarray}
\rho_{+}\frac{dX}{dt} &=& - \frac{\partial \rho_{R}(x,t)}{\partial x}\Big|_{x=X^+} 
\label{StefanR} \\
\rho_{-}\frac{dX}{dt} &=& - \frac{\partial \rho_{L}(x,t)}{\partial x}\Big|_{x=X^-} 
\label{StefanL} \\
r_{+} ( 1 - \rho_{+}) &=& r_{-} ( 1 - \rho_{-})
\label{MatchLR}
\end{eqnarray}
The Stefan-type boundary conditions Eq.~(\ref{StefanR}) and  Eq.~(\ref{StefanL}), result from  the conservation of the number of particles. Eq.~(\ref{MatchLR}) expresses the fact that the average speed of the tracer vanishes. 

Hence, we must solve two  diffusion equations \eqref{DE} in the regions $-\infty<x< X(t)$ and $X(t)<x<+\infty$ subject to the initial condition 
\beq
\rho(x,0) = \rho_{-\infty} \theta(-x) + \rho_{+\infty} \theta(x) \label{eq:rhoit}
\eeq 
with the boundary conditions \eqref{StefanR} and \eqref{StefanL}. The two solutions are matched using \eqref{MatchLR}.

The Stefan problem~(\ref{StefanR}) admits a scaling solution of the form
\begin{equation}
\rho_{R}(x,t) = f(\xi), \quad \mathrm{where} \qquad \xi=\frac{x}{X}
\end{equation}
With  this scaling form, the heat equation and the boundary condition~\eqref{StefanR} 
reduce  to the ordinary differential equations
\begin{eqnarray}
  f'' &=& -c_{R} \,  \xi \,  f'   \,\, \hbox{ with } \,\, c_{R} = -\frac{ f'(1)}{\rho_{+}}
\label{ScaledStefanR}  \\
 X\,\dot X &=&  c_{R} \label{MotionR}
\end{eqnarray}
where the prime (the dot) denotes the derivative with respect to the scaled distance $\xi$ (the time). Solving  with the condition $f(1) = \rho_{+}$ leads  us to 
\begin{eqnarray}
\rho_{R}(x,t) &=&  \rho_{+} \left( 1  -  c_{R} \int_1^{\frac{x}{X}} e^{ c_{R}(1 -\xi^2)/2 }d\xi   \right)
\label{SolutionR} \\
X_t &=& \sqrt{ 2 c_R t } \label{SolXR} 
\end{eqnarray}
The validity of the solution requires that $c_R>0$. Since $\rho_R$ is monotonically decreasing, this implies that $\rho_+>\rho_{+\infty}.$ In this case, the tracer moves in the positive direction.
\begin{eqnarray}
\rho_{L}(x,t) &=&  \rho_{-} \left( 1  +  c_{L} \int_{\frac{x}{X}}^1 e^{ c_{L}(1 -\xi^2)/2 }d\xi   \right)
\label{SolutionL} \\
X_t &=& \sqrt{ 2 c_L t } \label{SolXL}
\end{eqnarray}
Comparing  \eqref{SolXR} with  \eqref{SolXL} gives us $c_R = c_L$: this  common value will be denoted by $c$. Recalling that $\rho_R(+\infty)=\rho_{+\infty}$ and $\rho_R(-\infty)=\rho_{-\infty}$, we can also deduce the relations
\beqa
\frac{\rho_{+\infty}}{\rho_+} &=&  1  -  c_{R} \int_1^{+\infty} e^{ c_{R}(1 -\xi^2)/2 }d\xi  =  1 -  \sqrt{ 2 \pi c} \, \,  e^{ c/2} \, \, 
\frac{ 1 -  \hbox{erf}(\sqrt{\frac{c}{2}})}{2} \label{BC+infty}\\
\frac{\rho_{-\infty}}{\rho_-} &=& 1  + c_{L} \int_{-\infty}^1 e^{ c_{L}(1 -\xi^2)/2 }d\xi = 1 +  \sqrt{ 2 \pi c} \,\,   e^{ c/2} \,  \, 
\frac{ 1 +  \hbox{erf}(\sqrt{\frac{c}{2}})}{2} \label{BC-infty}
\eeqa
Equations \eqref{BC+infty}, \eqref{BC-infty} together with the matching condition $r_{+} ( 1 - \rho_{+}) = r_{-} ( 1 - \rho_{-})$ (see eqn. \eqref{MatchLR}), provide us with three independent relations for the three unknowns $\rho_{+}, \rho_{-}$ and $c$. Eqn. \eqref{MatchLR} shows that the absolute values of $r_{+}$ and $r_{-}$ do not matter, only the asymmetry ratio $ -1 < \alpha= \frac{r_{+}-r_{-}}{r_{+}+ r_{-}} < 1$ does. \footnote{A simple procedure to analyze these equations is as follows: for any given values of  $\rho_{\pm \infty}$, choose a  $c>0$ and this provides us with values of $\rho_{+}(c), \rho_{-}(c)$ and $\alpha(c)$.}

\begin{figure}
	\begin{subfigure}
		\centering
		\includegraphics[width=0.45\columnwidth]{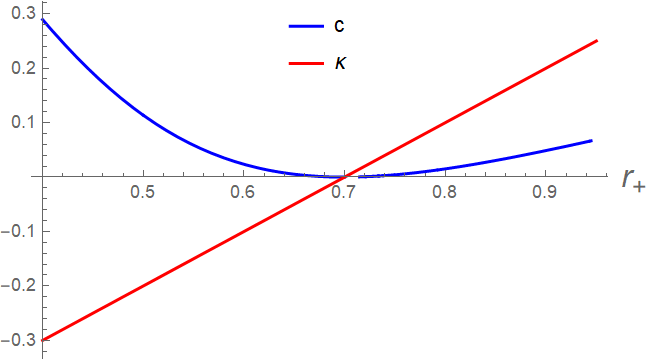}
	\end{subfigure}
	\begin{subfigure}
		\centering
		\includegraphics[width=0.45\columnwidth]{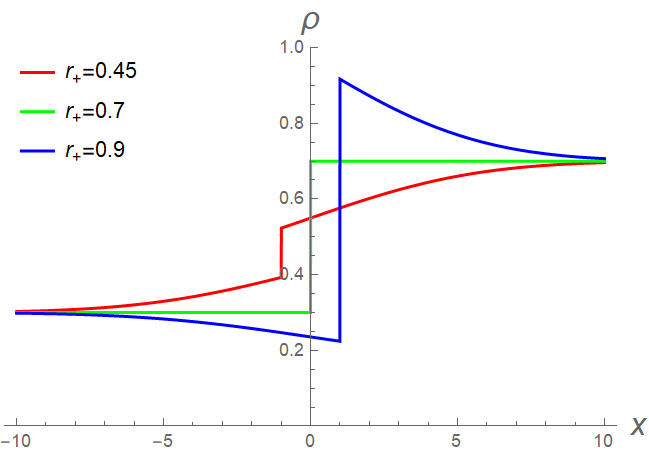}
	\end{subfigure}
	\caption{(a) A plot of $\kappa$ and $c$ as a function of $r_+$, on the hyperplane of parameters where $r_+ + r_- = 1$, for $\rho_{+\infty} = 0.7$ and $\rho_{-\infty}=0.3$. (b) A plot of the density profile for $t=10$ for three values of $r_+$, with the same values of other parameters as in (a). It can be seen that the particle remains stationary for $r_+=0.7$, and the density profile remains constant with time.}
	\label{drivensepplot}
\end{figure}

So far, we have assumed that $c>0$, and this implies that $\rho_+ \ge \rho_{+\infty}$ and $\rho_- \le \rho_{-\infty}$. Since eqn. \eqref{MatchLR} holds, we define the quantity
\beq
\kappa = r_{+} ( 1 - \rho_{+\infty}) - r_{-} ( 1 - \rho_{-\infty})
\eeq
It can be seen that $\kappa>0$ when $c>0$ and $\kappa=0$ when $c=0$. Thus, when
\beqa
r_{+} ( 1 - \rho_{+\infty}) = r_{-} ( 1 - \rho_{-\infty}), \mbox{  we have  } X_t = 0 \nonumber
\eeqa
and the tracer does not move. Now, if $\kappa < 0$, assuming for the moment that $r_+>r_-$, the bias is not large enough to overcome the push of the density gradient, and the particle moves in the negative direction. To obtain the solution for the case $\kappa<0$, we consider the mirrored system, where one exchanges $r_+$ and $r_-$, and exchanges $\rho_{+\infty}$ and $\rho_{-\infty}$. For this mirrored system, we have $\kappa>0$ and thus can apply the method above. Let $\rho_{\rm{mirror}}$ be the density profile for this mirrored system, and $c_{\rm{mirror}}$ the value of $c$. The solution for $\kappa<0$ is then
\beqa
\rho(x,t)&=& \rho_{\rm{mirror}}(-x,t) \nonumber \\
X_t &=& - \sqrt{2 c_{\rm{mirror}} t} \nonumber. 
\eeqa
Thus, we have that $X = \mbox{Sgn}(\kappa)\sqrt{2 c t}$. Fig. \ref{drivensepplot} shows a plot of the density profile around the tracer for various values of $\kappa$. 

One can look at the limit of a small asymmetry around the equilibrium state, $r_{\pm} = (1 \pm \epsilon)$, and $\rho_{+\infty}=\rho_{-\infty}=\rho$. In this limit,
\beqa
c &=& \frac{2}{\pi} \frac{(1-\rho)^2}{\rho^2} \epsilon^2 + O(\epsilon^3), \nonumber\\
X_t &=& \frac{2}{\sqrt{\pi}}\frac{(1-\rho)}{\rho}\epsilon \sqrt{t} + O(\epsilon^2) 
\eeqa
The value of the drift under a small bias is thus the same as the (annealed) variance without a bias, as expected from the Einstein relation \cite{ferrari1985diffusion,landim1998driven}.

In  the totally asymmetric case $r_{-}=0$, we have  $\rho_{+} =1$ irrespective of the value of $r_{+}$. Then, in  the limiting cases of small and large density, the average position of the tracer  exhibits the following asymptotic behaviors:
\begin{equation}
X_t \simeq  \sqrt{t}\times
\begin{cases}
\sqrt{2/\rho_{+\infty}}             & \rho_{+\infty}\downarrow 0\\
2\pi^{-1/2}(1-\rho_{+\infty})   & \rho_{+\infty}\uparrow 1
\end{cases}
\end{equation}

\subsection{Boundary conditions at the origin}

In the SEP, a biased tracer attempts a hop to the right with probability $r_+$, and to the left with probability $r_-$. If $r_+=r_-=1$, the tracer is unbiased and performs the same motion as all other particles. We now consider a driven tracer in a general single-file system and define the biased dynamics carefully. If an unbiased tracer in a configuration $C$ has a probability $p_{\rm{T}}(C)$ of hopping to the right or the left, we define the driven tracer to hop to the right with rate $r_+ p_{\rm{T}}(C)$ and to the left with $r_- p_{\rm{T}}(C)$. 

In the mass-transfer picture, the movement of the tracer corresponds to biased mass transfer across the bond $[-1,0]$. The mass transfer across this bond to the left is enhanced by a factor $r_+$ while transfer to the right is enhanced (or suppressed) by a factor $r_-$. The total displacement of the tracer is, in the ZRP picture, is minus the total current through the origin from time $0$ to time $t$, using eqn. \eqref{transf:trace}.

We will consider the MFT solutions on the two sides of the origin as separate systems on the two half-planes \cite{meerson2014large}, whose solutions are related through certain boundary conditions at the origin. Assuming that the coarse-grained hydrodynamic limit remains valid, the hydrodynamic local density field during the evolution is then given by $q(x,t)$, the solution of the MFT equations. This implies that local equilibrium holds, with the local density parameter given by $q([\ell x],t)$ where $\ell$ is the coarse-graining length. We also assume that the joint distribution on different sites factorizes \cite{landim1998driven,landim2000equilibrium}. For local equilibrium to hold on both sides,
\beq
r_- R(\rho_{-1}) = r_+ R(\rho_{0}) \nonumber
\eeq
where $R(\rho)$ denotes the total rate of outward mass transfer from a site in the steady state with density $\rho$ (we have assumed local equilibrium). Thus, in the hydrodynamic limit,
\beq
r_- R(q(0^-,t)) = r_+ R(q(0^+,t)), \mbox{  or  }  r_- R(q_-) = r_+ R(q_+) \label{eq:qcond}
\eeq
where we have used the notation $g_+ \equiv g(x=0^+)$ and $g_- \equiv g(x=0^-)$ for the limits of a function $g$ on either sides of the origin. The function $R$ is related to the diffusivity as, in the presence of a uniform density gradient, the current in the continuum limit is given by
\beq
j = -\frac{1}{2}\p_x R(\rho), \mbox{  and hence,  } R'(\rho) = 2 D(\rho) \label{eq:foccD}
\eeq
where $D(\rho)$ is the diffusivity. For the ZRP, we have $R(q) = \frac{2 q}{1+q}$, and the condition eqn. \eqref{eq:qcond} becomes
\beq
r_- \frac{q_-}{1+q_-} = r_+ \frac{q_+}{1+q_+} \label{eq:qcondzrp}
\eeq
It is seen that this is equivalent to the SEP condition for the density profile, eqn. \eqref{MatchLR}.

We now find the boundary conditions for $p$ across the origin by considering the variation of the action. The action for a trajectory is given by \cite{Bertini2015}
\beq
S = \int_0^T dt \int_{-\infty}^{\infty} dx \left( p \p_t q - \frac{\sigma(q)}{2} (\p_x p)^2 + D(q) (\p_x p) (\p_x q) \right) \label{eq:MFTact}
\eeq

To derive the boundary conditions on $p$ at the origin, we consider the variation of the action with respect to $q$ and $p$. (See the  \ref{app:boundaryp} for the details of the derivation.) First, we consider the variation with respect to $q$, which gives boundary conditions relating the gradient of $p$ on the two sides of the origin:
\beq
r_- (\p_x p)_+ = r_+ (\p_x p)_- \label{eq:pcond1}
\eeq

Similarly, we can consider the variation of the action with respect to $p$. This gives boundary conditions that relate $p$ on the two sides of the origin,
\beq
p_+ = p_-\label{eq:pcond2}
\eeq

Eqns. \eqref{eq:qcond}, \eqref{eq:pcond1} and \eqref{eq:pcond2} complete the description of boundary conditions at the origin for a biased particle. It can be seen that the values of $r_+$ and $r_-$ come into the boundary conditions only upto an overall multiplicative constant, and hence only the value of $\frac{r_+}{r_++r_-}$ matters to the dynamics.

Since the current at the origin is related to the displacement of the tracer, we have
\beq
\frac{dX}{dt} = -j_m\vert_{x=0^+} = -j_m\vert_{x=0^-} \label{eq:tracerJ}
\eeq


Eqn. \eqref{eq:qcondzrp} becomes, in the SEP MFT,
\beq
r_+ (1-q_s)\vert_{x=X^+} = r_- (1-q_s)\vert_{x=X^-}
\eeq
The two equations above are the MFT generalization of the conditions in eqns. \eqref{StefanR}, \eqref{StefanL} and \eqref{MatchLR}. Based on these MFT equations, we propose the following fluctuating hydrodynamic equations to model the tracer hydrodynamics. In the bulk on either side of the tracer we have the equations
\beqa
\p_t \rho_s &=& - \p_x j_s, \label{eq:tracerhydro1} \\
j_s &=& -D(\rho_s) \p_x \rho_s + \sqrt{\sigma_s(\rho_s)} \eta
\eeqa
where $\eta$ is a noise uncorrelated in space and time. These bulk equations are supplemented by the boundary conditions at the tracer:
\beqa
\frac{dX}{dt} &=& \left(\frac{1}{\rho_s} j_s \right) \bigg\vert_{x=X^+} = \left(\frac{1}{\rho_s} j_s \right) \bigg\vert_{x=X^-} \\
r_+ (1-\rho_s)_+ &=& r_- (1-\rho_s)_- \label{eq:tracerhydro4}
\eeqa
where for the first equation we have used eqns. \eqref{eq:Jmap} and \eqref{eq:tracerJ}.

\section{Fluctuations of the  driven tracer in the high density limit} \label{sec:highdensity}


We now consider the low-density limit in the ZRP, which corresponds to the high-density limit of the SEP. In this limit, the hydrodynamic coefficients for the ZRP become
\beq
D(q) = 1 + O(q), ~~~ \sigma(q) = 2 q + O(q^2) 
\eeq

Away from the tracer particle, the bulk MFT equations for the ZRP hold,
\begin{subequations}
\beqa
\p_t q &=& \p_x( (\p_x q) - 2 q (\p_x p) ) \label{eq:ELZRP1} \\
\p_t p &=& - \p_x^2 p - (\p_x p)^2 \label{eq:ELZRP2}
\eeqa
\end{subequations}

The Cole-Hopf transformation
\beqa
Q = q e^{-p}, ~~~~ P = e^p 
\eeqa
simplifies the E-L equations into pure diffusion equations

\beq
\p_t Q = \p_x^2 Q, ~~~~ \p_t P = - \p_x^2 P \label{eq:Qeq}
\eeq

In the ZRP, in the high-density limit, the SEP initial condition \eqref{eq:rhoit}, the initial condition for the average density becomes
\beq
\rho_m(x,0) = (1-\rho_{+\infty}) \theta(x) + (1-\rho_{-\infty}) \theta(-x) + O((1-\rho)^2)
\eeq

The boundary condition on $p(x,T)$, eqn. \eqref{eq:pft} becomes
\beq
P(x,T) = 1 + (e^{\lambda}-1) \theta(x) \label{eq:Pft}
\eeq
The quenched initial condition \eqref{eq:qit} gives
\beq
Q(x,0) = \frac{1}{P(x,0)} \left( (1-\rho_{-\infty}) \theta(-x) + (1-\rho_{+\infty}) \theta(x) \right) \label{eq:Qqit}
\eeq

For the annealed initial condition, we have, from \eqref{eq:pit} with $D(q)=1$ and $\sigma(q) = 2 q$,
\beqa
q(x,0) &=& (1-\rho_{-\infty}) e^{p(x,0)} \theta(-x) + (1-\rho_{+\infty}) e^{p(x,0)-\lambda} \theta(x) \label{eq:sepannic}, \mbox{ giving,}\\
Q(x,0) &=&  (1-\rho_{-\infty}) \theta(-x) + (1-\rho_{+\infty}) e^{-\lambda} \theta(x) \label{eq:Qait}.
\eeqa

\subsection{Analytic solution for $P(x,t)$}

$P(x,t)$ obeys the anti-diffusion equation
\beq
\p_t P = - \p_x^2 P
\eeq
with initial condition
\beq
P(x,T) = 1 + (e^{\lambda}-1) \theta(x) \nonumber
\eeq
and boundary conditions
\beqa
P(x,t) &\rightarrow& e^{\lambda} \mbox{ as } x \rightarrow \infty, \nonumber\\
P(x,t) &\rightarrow& 1 \mbox{ as } x \rightarrow -\infty \nonumber
\eeqa
The general solution is thus
\beqa
P(x,t) = 
\begin{cases}
	&e^{\lambda} + A \left(1 - \mbox{erf}\left(\frac{x}{\sqrt{4(T-t)}} \right) \right) \mbox{  for  } x>0 \\
	&1 + B \left(1 + \mbox{erf}\left(\frac{x}{\sqrt{4(T-t)}} \right) \right) \mbox{  for  } x<0
\end{cases}
\eeqa
where $A$ and $B$ are constants to be determined from the boundary conditions. We now use the two conditions \eqref{eq:pcond1} and \eqref{eq:pcond2}
\beq
P_+ = P_-, \mbox{  and  } r_- (\p_x P)_+ = r_+ (\p_x P)_- \nonumber
\eeq
to solve for $A$ and $B$. This gives, 
\beqa
A = \frac{r_+}{r_+ + r_-} (1-e^{\lambda}), \mbox{ and } B = \frac{r_-}{r_+ + r_-} (e^{\lambda}-1) \nonumber 
\eeqa

Thus, we have
\beqa
P(x,t) = 
\begin{cases}
	&e^{\lambda} + \frac{r_+}{r_+ + r_-} (1-e^{\lambda}) \mbox{erfc}\left(\frac{x}{\sqrt{4(T-t)}} \right) \mbox{  for  } x>0 \\
	&1 + \frac{r_-}{r_+ + r_-} (e^{\lambda}-1) \mbox{erfc}\left(\frac{|x|}{\sqrt{4(T-t)}} \right) \mbox{  for  } x<0
\end{cases} \label{eq:Psol}
\eeqa


We want to determine the cumulant generating function (CGF) of the position of the tagged particle after time $T$,
\beq
\mu^p(\lambda) = \log{\big\la e^{\lambda X_T} \big\ra}
\eeq

As the position of the tagged particle in the SEP is related to the current in the ZRP through \eqref{transf:trace}, we have the relation
\beq
\mu^p(\lambda) = \log{\big\la e^{\lambda X_T} \big\ra} = \log{\big\la e^{-\lambda J(T)} \big\ra} = \mu(-\lambda)
\eeq
where $\mu(\lambda)$ is the CGF of the current in the ZRP, defined in \eqref{eq:mudef}. Now, using \eqref{eq:mupformula} we can calculate $\mu(\lambda)$ through the MFT solution, 
\beqa
\mu'(\lambda) &=& \int_0^{\infty} dx (q(x,T) - q(x,0)) \nonumber \\
&=& \int_0^{\infty} dx \left( Q(x,T) P(x,T) - Q(x,0) P(x,0) \right) \label{eq:mupmain}
\eeqa

$Q(x,t)$ evolves according to a pure diffusion equation with the boundary condition at the origin, using eqn. \eqref{eq:qcondzrp},\\
\beq
r_+ Q(0^+,t) = r_- Q(0^-,t) \label{eq:Qcond}
\eeq

Conservation of the current of $q$ across the origin gives
\beq
(\p_x Q)_- = (\p_x Q)_+
\eeq

Consider a general initial condition
\beq
Q(x,0) = Q_1(x) \Theta(x) + Q_2(x) \Theta(-x)
\eeq
The general solution for $x>0$ satisfying the boundary conditions is (see \ref{app:Qsol} for details)
\beqa
Q(x,t) = \int_0^{\infty} \frac{dy}{\sqrt{4\pi t}} &\bigg[& Q_1(y) \left( e^{-\frac{(x-y)^2}{4t}} - \frac{r_+ - r_-}{r_+ + r_-} e^{-\frac{(x+y)^2}{4t}} \right) \nonumber\\
& &+ \frac{2 r_-}{r_+ + r_-} Q_2(y) e^{-\frac{(x+y)^2}{4t}} \bigg] \nonumber
\eeqa

Inserting the solutions for $Q$ and $P$ into eqn. \eqref{eq:mupmain}, and performing the integral over $x$, we get
\beqa
\mu'(\lambda) &=& -\int_0^{\infty} dy Q_1(y) \left( e^{\lambda} + \frac{r_+}{r_+ + r_-} (1-e^{\lambda}) \mbox{erfc}\left(\frac{y}{\sqrt{4T}} \right) \right) \nonumber\\
 & & + \int_0^{\infty} dy Q_1(y) e^{\lambda} \left(1 - \frac{r_+}{r_+ + r_-} \mbox{erfc}\left(\frac{y}{\sqrt{4T}} \right) \right)  \nonumber\\
 & & + \int_0^{\infty} dy Q_2(y) e^{\lambda} \frac{r_-}{r_+ + r_-} \mbox{erfc}\left(\frac{y}{\sqrt{4T}} \right) \nonumber \\
 &=& \int_0^{\infty} dy \left( -\frac{r_+}{r_+ + r_-} Q_1(y) \mbox{erfc}\left(\frac{y}{\sqrt{4T}} \right) + \frac{r_-}{r_+ + r_-} Q_2(y) e^{\lambda} \mbox{erfc}\left(\frac{y}{\sqrt{4T}} \right) \right) \nonumber\\
 \label{eq:mup1}
\eeqa

\subsection{Cumulants  in the Annealed Case}

For the annealed case, inserting the initial condition \eqref{eq:Qait} into eqn. \eqref{eq:mup1},

\beqa
\mu_a'(\lambda) &=& \frac{\sqrt{4T}}{r_+ + r_-} \left(-(1-\rho_{+\infty}) r_+ e^{-\lambda} + (1-\rho_{-\infty}) r_-  e^{\lambda}\right) \int_0^{\infty} dy~ \mbox{erfc}(y) \nonumber\\
&=& \sqrt{\frac{4T}{\pi}}\frac{1}{r_+ + r_-} \left(-(1-\rho_{+\infty}) r_+ e^{-\lambda} + (1-\rho_{-\infty}) r_-  e^{\lambda}\right)
\eeqa

We thus have, using the condition that $\mu_a(0) = 0$,
\beq
\frac{\mu^{p}_a(\lambda)}{\sqrt{4T}} = \frac{\mu_a(-\lambda)}{\sqrt{4T}} =
\frac{(1-\rho_{+\infty}) r_+ (e^{\lambda}-1) + (1-\rho_{-\infty}) r_- (e^{-\lambda}-1)}{\sqrt{\pi}(r_+ + r_-)} + O((1-\rho)^2) \label{eq:muafinal}
\eeq
where we have made explicit that the results here are derived in the high-density limit of the SEP.

\subsection{Cumulants in the Quenched Case}

For the quenched case, we have the initial condition \eqref{eq:Qqit}, which gives,
\beq
\mu_q'(\lambda) = \int_0^{\infty} dy \left(-\frac{ (1-\rho_{+\infty}) \frac{r_+}{r_+ + r_-} e^{-\lambda} \mbox{erfc}\left(\frac{y}{\sqrt{4 T}}\right)}{1 + \frac{r_+}{r_+ + r_-} (e^{-\lambda}-1) \mbox{erfc}\left(\frac{y}{\sqrt{4T}}\right)} + \frac{(1-\rho_{-\infty}) \frac{r_-}{r_+ + r_-} e^{\lambda} \mbox{erfc}\left(\frac{y}{\sqrt{4 T}}\right)}{1 + \frac{r_-}{r_+ + r_-} (e^{\lambda}-1) \mbox{erfc}\left(\frac{y}{\sqrt{4T}} \right)} \right)
\nonumber 
\eeq

Which then gives, for the CGF,
\beqa
\frac{\mu^p_q(\lambda)}{\sqrt{4T}} &=& \frac{\mu_q(-\lambda)}{\sqrt{4T}} = \int_0^{\infty} dy \bigg[ (1-\rho_{+\infty}) \log{\left(1 + \frac{r_+}{r_+ + r_-} (e^{\lambda}-1) \mbox{erfc}(y) \right)} \nonumber\\
& &+ (1-\rho_{-\infty}) \log{\left(1 + \frac{r_-}{r_+ + r_-} (e^{-\lambda}-1) \mbox{erfc}(y) \right)} \bigg] + O((1-\rho)^2) \nonumber\\
\label{eq:muqfinal}
\eeqa
where we have made the high-density limit explicit. Equations \eqref{eq:muafinal} and \eqref{eq:muqfinal} are in agreement with the results of \cite{Jussieu13,poncet2021cumulant}, derived using a microscopic approach.

\subsection{The variance for a general initial condition}

Following Banerjee et al \cite{banerjee2022role}, we now consider more general initial conditions described by a generalized susceptibility, such that the log-probability of a small density variation in the initial ensemble is given by
\beq
F(\{q(x,0)\}) \approx \frac{1}{2} \int_{-\infty}^{0} dx \frac{(q(x,0)-(1-\rho_{-\infty}))^2}{\alpha_{\rm{ic}} (1-\rho_{-\infty})} + \frac{1}{2} \int_{0}^{\infty} dx \frac{(q(x,0)-(1-\rho_{+\infty}))^2}{\alpha_{\rm{ic}} (1-\rho_{+\infty})} 
\eeq
where we have kept terms only to quadratic order in the variation. The quantity $\alpha_{\rm{ic}}$ defines a generalized susceptibility, and is related to the Fano factor \cite{banerjee2022role}. The generalized susceptibility $\alpha_{\rm{ic}}$ describes the amount of fluctuations in the initial ensemble. For the quenched or hyperuniform \cite{torquato2003local,dandekar2020exact} initial conditions, $\alpha_{\rm{ic}} = 0$, while for annealed initial conditions, because local equilibrium is a Poisson state, $\alpha_{\rm{ic}} = 1$.

We analyze the dependence of the variance of the biased tracer position on $\alpha_{\rm{ic}}$. This involves terms to second order in the CGF $\mu(\lambda)$, and hence to first order in $q(x,t)$. For an ensemble of initial conditions described by the log-probability function $F$, the initial condition for $q$ is given implicitly by \cite{derrida2009current}
\beq
p(x,0) = \lambda \theta(x) + \frac{\delta F}{\delta q(x,0)},
\eeq 
which gives
\beq
q(x,0) \approx \theta(-x) (1-\rho_{-\infty}) \big( 1+ \alpha_{\rm{ic}} p(x,0) \big) + \theta(x) (1-\rho_{+\infty}) \big( 1 + \alpha_{\rm{ic}} (p(x,0) - \lambda) \big)
\eeq
to $O(\lambda^2)$ and $O((1-\rho)^2)$. For $\alpha_{\rm{ic}}=0$, we have the quenched initial conditions, while the case $\alpha_{\rm{ic}}=1$ is equivalent to eqn. \eqref{eq:sepannic} first order in $\lambda$. Since $q(x,t)$ is a linear function of $\alpha_{\rm{ic}}$, and the MFT equations \eqref{eq:EL1} and \eqref{eq:EL2} are linear to first order in $\lambda$ \cite{krapivsky2012fluctuations}, it is expected that $\mu'(\lambda)$ to first order for  general $\alpha_{\rm{ic}}$ will be a linear combination of the quenched and annealed answers.

Using the Cole-Hopf transformation and eqn. \eqref{eq:Psol}, and keeping terms only to $O(\lambda)$ and in the high-density limit,
\beqa
Q_1(x) &=& (1-\rho_{+\infty}) - \lambda ~(1-\rho_{+\infty}) (1-\alpha_{\rm{ic}}) \left(1-\frac{r_+}{r_+ + r_-} \mbox{Erfc}\left(\frac{y}{\sqrt{4 T}}\right) \right) \nonumber\\
Q_2(x) &=& (1-\rho_{-\infty}) - \lambda ~\frac{r_-}{r_+ + r_-} (1-\rho_{-\infty}) (1-\alpha_{\rm{ic}}) \mbox{Erfc}\left(\frac{y}{\sqrt{4 T}}\right) \nonumber
\eeqa

Inserting into eqn. \eqref{eq:mup1}, we have
\beqa
\mu'(\lambda) &=& \frac{\sqrt{4T}}{\sqrt{\pi} (r_+ + r_-)} \bigg\{ (1-\rho_{-\infty})r_- -(1-\rho_{+\infty}) r_+ \nonumber\\
& &+\lambda\bigg[(1-\rho_{+\infty}) r_+ \left(1-\left(\sqrt{2}-2\right) (\alpha_{\rm{ic}}-1)\frac{r_+}{r_+ + r_-}\right) \nonumber \\
& & +(1-\rho_{-\infty}) r_- \left(1-\left(\sqrt{2}-2\right) (\alpha_{\rm{ic}}-1) \frac{r_-}{r_+ + r_-}\right)\bigg]\bigg\} \nonumber\\ 
&=& J_{\rm{quenched}} + \alpha_{\rm{ic}} \left( J_{\rm{annealed}} - J_{\rm{quenched}} \right) + O(\lambda^2) + O((1-\rho)^2) \label{eq:mupalpha}
\eeqa
The cases $\alpha_{\rm{ic}}=0$ and $\alpha_{\rm{ic}}=1$ give back the results, to $O(\lambda)$, of eqns. \eqref{eq:muqfinal} and \eqref{eq:muafinal}. From eqn. \eqref{eq:mupalpha}, we get that the second cumulant of the biased tracer position in the high-density limit is
\beq
\la X^2 \ra_c = \la X^2 \ra_{c,\rm{quenched}} + \alpha_{\rm{ic}} \left( \la X^2 \ra_{c,\rm{annealed}} - \la X^2 \ra_{c,\rm{quenched}} \right)
\eeq
Thus, the dependence of the second cumulant on $\alpha_{\rm{ic}}$ follows the same pattern for a biased tracer as that found for unbiased tracers in \cite{banerjee2022role}.

\section{Concluding remarks} \label{sec:conclusions}

In this paper, we have investigated the problem of a biased tracer in a single-file system.  Thanks
to a mapping to a ZRP with a biased bond at the origin, we could write exact
boundary conditions at the origin within the MFT framework, solve  the high density limit, retrieve at the macroscopic scale the results of microscopic
calculations  \cite{Jussieu13,poncet2021cumulant}
and study a whole family  of initial conditions, thus 
generalizing the results of \cite{banerjee2022role} to biased tracers.

We emphasize that the  boundary conditions derived here are valid for general single-file systems and thus open up the MFT framework for investigating biased tracers. The fluctuating hydrodynamic equations \eqref{eq:tracerhydro1}- \eqref{eq:tracerhydro4} could also be useful for numerical investigation
as well as for perturbative calculations. We expect that this
continuous  approach,  that takes fluctuations into account, will be useful
for investigating the effect of a local defect  at the global 
hydrodynamic scale.

\vskip 1cm

{\it While completing this manuscript, we became aware of a related work
  `Duality in single-file diffusion' by  our collegues P. Rizkallah, A.  Grabsch, P.  Illien and  Olivier B\'enichou from  Sorbonne University
  (arXiv 2207.07549): they   also
  consider the  mapping
  from a single-file  to a mass-transfer process to 
 find  relations between various models and  to transfer
  exact solutions between them. We were
  motivated by a different problem, the biased tracer and 
  the mapping was introduced 
  to rewrite a (stochastic)  moving Stefan  problem as a
  localized boundary condition.}

\section*{Acknowledgements}

We would like to thank Paul Krapivsky for early collaboration and helpful discussions and comments on the manuscript.
  
\appendix

\section{Boundary conditions on $p$ at the biased particle} \label{app:boundaryp}

First we consider the variation in the MFT action $S$ (eqn. \eqref{eq:MFTact}) in terms of the variation in $q$. We only keep the terms which are total derivatives in space (the bulk variation gives the E-L equations \eqref{eq:EL1} and \eqref{eq:EL2} and the total time-derivative terms give the boundary conditions \cite{derrida2009current}),
\beqa
\delta S &=& \int_0^T dt \bigg(\big[\delta q D(q) \p_x p\big]_{x=\infty} -  \big[\delta q D(q) \p_x p\big]_{x=0^+} \nonumber\\
&& + \big[\delta q D(q) \p_x p\big]_{x=0^-} - \big[\delta q D(q) \p_x p\big]_{x=-\infty} \bigg) \label{eq:Svarq}
\eeqa

We now use the fact that $(\p_x p) \rightarrow 0$ as $x \rightarrow \pm \infty$, to get
\beqa
0 &=& \int_0^T dt \left(\big[\delta q_+ D(q_+)\p_x p\big]_+ - \big[\delta q_- D(q_-)\p_x p\big]_-\right) \nonumber\\
 &=& \int_0^T dt \left(\delta R(q_+) (\p_x p)_+ - \delta R(q_-) (\p_x p)_- \right) \nonumber
\eeqa
where we have used eqn. \eqref{eq:foccD}. Now using eqn. \eqref{eq:qcond}, we get that
\beq
r_+ \delta R(q_+) = r_- \delta R(q_-) \nonumber
\eeq
and hence,
\beq
\delta S = \int_0^T dt ~ \delta R(q_+) \left( r_- (\p_x p)_+ - r_+ (\p_x p)_- \right) =0 \nonumber
\eeq
and hence,
\beq
r_- (\p_x p)_+ = r_+ (\p_x p)_- 
\eeq

The boundary terms in the variation in $p$ can be similarly computed to be
\beq
\delta S = \int_0^T dt \left( \delta p_+ ( \sigma(q) (\p_x p) - D(q) (\p_x q) )_+ - \delta p_- ( \sigma(q) (\p_x p) - D(q) (\p_x q) )_- \right) \nonumber
\eeq

Now, it is known that in the MFT, the density field $q$ is conserved, and hence, the current flowing from the right to the origin is equal to the current flowing to the left from the origin,
\beq
( \sigma(q) (\p_x p) - D(q) (\p_x q) )_+ =  ( \sigma(q) (\p_x p) - D(q) (\p_x q) )_- \label{eq:qconserve}
\eeq

For this to happen, it must be that case that $\delta p_+$ and $\delta p_-$ cannot vary independently, that is,
\beq
\delta p_- = \delta p_+, \mbox{   that is,  } p_+ = p_- + C
\eeq

The constant $C$ can be fixed by looking at the right hand side of \eqref{eq:EL2}, where $C \neq 0$ would lead to an unphysical delta-function-squared term. Hence,
\beq
p_+ = p_- 
\eeq

\section{General solution for $Q(x,t)$} \label{app:Qsol}

In this Appendix we calculate a general formula for $Q(x,t) = q(x,t) e^{-p(x,t)}$ evolving according to the diffusion equation \eqref{eq:Qeq} and the boundary conditions
\beq
r_+ Q(0^+,t) = r_- Q(0^-,t) \mbox{ for $t>0$, and } (\p_x Q)_+ = (\p_x Q)_+ \label{eq:Qbounds}
\eeq
The latter equation comes from the conservation of the current of $q$ across the origin, \eqref{eq:qconserve} and using the boundary conditions \eqref{eq:qcondzrp}, \eqref{eq:pcond1} and \eqref{eq:pcond2}. We consider the initial condition
\beqa
Q(x,0) = Q_1(x) \Theta(x) + Q_2(x) \Theta(-x) \nonumber
\eeqa

Since the values of $Q$ on either side of the origin are constrained to be different, we consider a partial mirror-image solution on the two sides of the origin,
\beqa
Q(x,0) = 
\begin{cases}
	\int_0^{\infty} \frac{dy}{\sqrt{4\pi t}} \left( Q_1(y) \left( e^{-\frac{(x-y)^2}{4t}} - a e^{-\frac{(x+y)^2}{4t}} \right) + (1-a) Q_2(y) e^{-\frac{(x+y)^2}{4t}} \right) \\
	\mbox{ for $x>0$}\\
	\int_{-\infty}^{0} \frac{dy}{\sqrt{4\pi t}} \left( Q_2(y) \left( e^{-\frac{(x-y)^2}{4t}} - b e^{-\frac{(x+y)^2}{4t}} \right) + (1-b) Q_1(y) e^{-\frac{(x+y)^2}{4t}}\right) \\
	\mbox{ for $x<0$}
\end{cases}
\nonumber
\eeqa
where $a$ and $b$ are constants determined from the boundary conditions in \eqref{eq:Qbounds}, which give
\beq
r_+ (1-a) = r_- (1-b) \mbox{  and  } a = -b, \nonumber
\eeq
Hence, we have
\beq
a = -b = \frac{r_+ - r_-}{r_+ + r_-}
\eeq

\section*{References}

\bibliographystyle{unsrt}
\bibliography{tracer}

\end{document}